# *Operando* Electrochemical Kinetics in Particulate Porous Electrodes by Quantifying the Mesoscale Spatiotemporal Heterogeneities

*Shubham Agrawal[1] and Peng Bai[1,2,]\**

[1]Department of Energy, Environmental & Chemical Engineering, Washington University in St. Louis, St. Louis, Missouri – 63130, United States

[2]Institue of Materials Science and Engineering, Washington University in St. Louis, St. Louis, Missouri – 63130, United States

*Email: pbai@wustl.edu

**Abstract**

Electrochemical energy systems rely on particulate porous electrodes to store or convert energies. While the three-dimensional porous structures were introduced to maximize the interfacial area for better overall performance of the system, spatiotemporal heterogeneities arose from materials thermodynamics localize the charge transfer processes onto a limited portion of the available interfaces. Here, we demonstrate a simple but precision method that can directly track and analyze the *operando* (i.e. local and reacting) interfaces at the mesoscale in a practical graphite porous electrode to obtain the true local current density, which turned out to be two orders of magnitude higher than the globally averaged current density adopted by existing studies. Our results sheds light on the long-standing discrepancies in kinetics parameters derived from electroanalytical measurements and from first principles predictions. Contradictory to prevailing beliefs, the electrochemical dynamics is not controlled by the solid-state diffusion process once the spatiotemporal reaction heterogeneities emerge.

# 1. Introduction

Electrochemical energy storage and conversion systems are critical for a sustainable future[1]. Lithium-ion batteries (LIBs) that offer the highest energy density have revolutionized electronic devices, portable power tools and electric cars.[2–4] But their further advancements have been impaired by the random occurrences of elusive safety accidents,[5,6] which are believed to originate from microscopic heterogeneities in the particulate porous electrodes.[7] State-of-charge (SOC) heterogeneities have recently been identified in both the solid-solution[8–12] and phase-transforming electrodes,[13–19] as a direct result of non-uniform distribution of electrochemical reactions due to either the structures of the composite porous electrodes[8,20] or the thermodynamics of the active materials.[21,22] While the nanoscale heterogeneities in individual particles detected by synchrotron X-ray provide deep insights on the possible degradation mechanisms, the evolutions of the heterogeneities among hundreds of particles sitting in realistic surroundings are critical for the understanding of the true local electrochemical kinetics that dictate the real-time performance.

The immediate consequence of the spatiotemporal heterogeneities is that the actual reacting interfacial area at any instant, i.e. area of the *operando* (local and working) electrochemical interface, is only a small portion of the total available interfacial area usually obtained from the Brunauer-Emmett-Teller (BET) method. Given that existing electroanalytical techniques[23,24] rely on the square-law scaling $D_{Li} \sim [I(t)/S]^2$ to extract the diffusion coefficient $D_{Li}$ from the total current $I(t)$ and the assumed *constant total* interfacial area $S$, the electrochemical kinetics in systems with strong heterogeneities may have been misinterpreted due to the smaller *operando* interfacial area.

As one of the most widely used electrodes for both the nonaqueous[15,25] and aqueous[26] batteries, graphite electrodes[27] are known to have strong reaction heterogeneities[13,15] reflected by its particle-by-particle reaction mechanism.[15,21] Depending on the choices of electrode area, e.g. BET or geometric, the lithium-ion diffusion coefficient in graphite ($D_{Li}$) extracted by the classic electroanalytical methods varies by about 8 orders of magnitude in the literature.[28–36] Still, $D_{Li}$ obtained for SOC ranges with phase transformation were always about 2 orders of magnitude lower than the average.[29,35] The discrepancy has long been doubted as the inaccuracy of the interfacial area,[37] but conclusive evidence is still missing. Similar orders-of-magnitude discrepancies also exist in other porous electrodes composed of phase-transforming[24] or solid-solution particles,[38–40] missing satisfactory explanations. The discrepancies in the kinetic parameters directly affect the determination of the rate-limiting step, and thereafter the validity of traditional electroanalytical techniques and the effectiveness of the predicted rational design strategies.

Here, we use graphite as a model system to demonstrate the direct quantification of the true local current densities for precision electrochemical kinetics. The unique color changing property during graphite lithiation[15,36] allows us to develop economical *operando* platform with optical microscope (**Figure S1**, Supporting Information) to investigate the dynamics of the heterogeneities at high speed and at the mesoscale (imaging hundreds of particles simultaneously every two seconds). Our study reveals that the state of charge (SOC) heterogeneities are indeed the result of reaction heterogeneities, which lead to the localization of the reaction flux onto a limited number of particles in the electrode. Using the moving phase boundaries between different stages (phases) of lithiated graphite to approximate the *operando* electrochemical interfacial area, the true local current density was determined to be as least 2 orders of magnitude higher than the averaged current density over the adjusted BET surface area. The insights gathered from the interface area and the true local current density suggest that, once the heterogeneities emerge, the *operando* (i.e. local and working) electrochemical kinetics of the porous electrode is not diffusion limited. This study highlights the need of tracking the phase boundaries to resolve the long-standing huge discrepancies between experiments and theoretical predictions.

## 2. Results

### 2.1. Spatiotemporal heterogeneities

We conducted three sets of potentiostatic intermittent titration technique (PITT) experiment with 10 mV, 100 mV and 200 mV steps, respectively. **Figure 1 (a)-(c)** demonstrates the color evolution in the thin graphite electrode (**Figure S2**, Supporting Information) during the 10 mV PITT experiment, with a threshold current of C/20. The entire lithiation process can be divided into three segments based on the colors of the lithiated graphite (Video S1 and **Figure S3**). In segment (I) all empty particles (dark grey) reacted concurrently regardless of their morphology and size to become blue (Stage 3). At this point, the blue particles accommodated 23% of the total capacity supplied in the entire PITT discharge and brought down the cell voltage from 275 mV to 85 mV. Since, at this moment (t = 0 s shown in Figure 1 (a)), all the particles were in Stage 3 (blue), the SOC associated with Stage 3 was determined to be the global SOC of the electrode, i.e., $x_B = 23\%$, slightly higher than the values adopted in earlier works.[15,41]

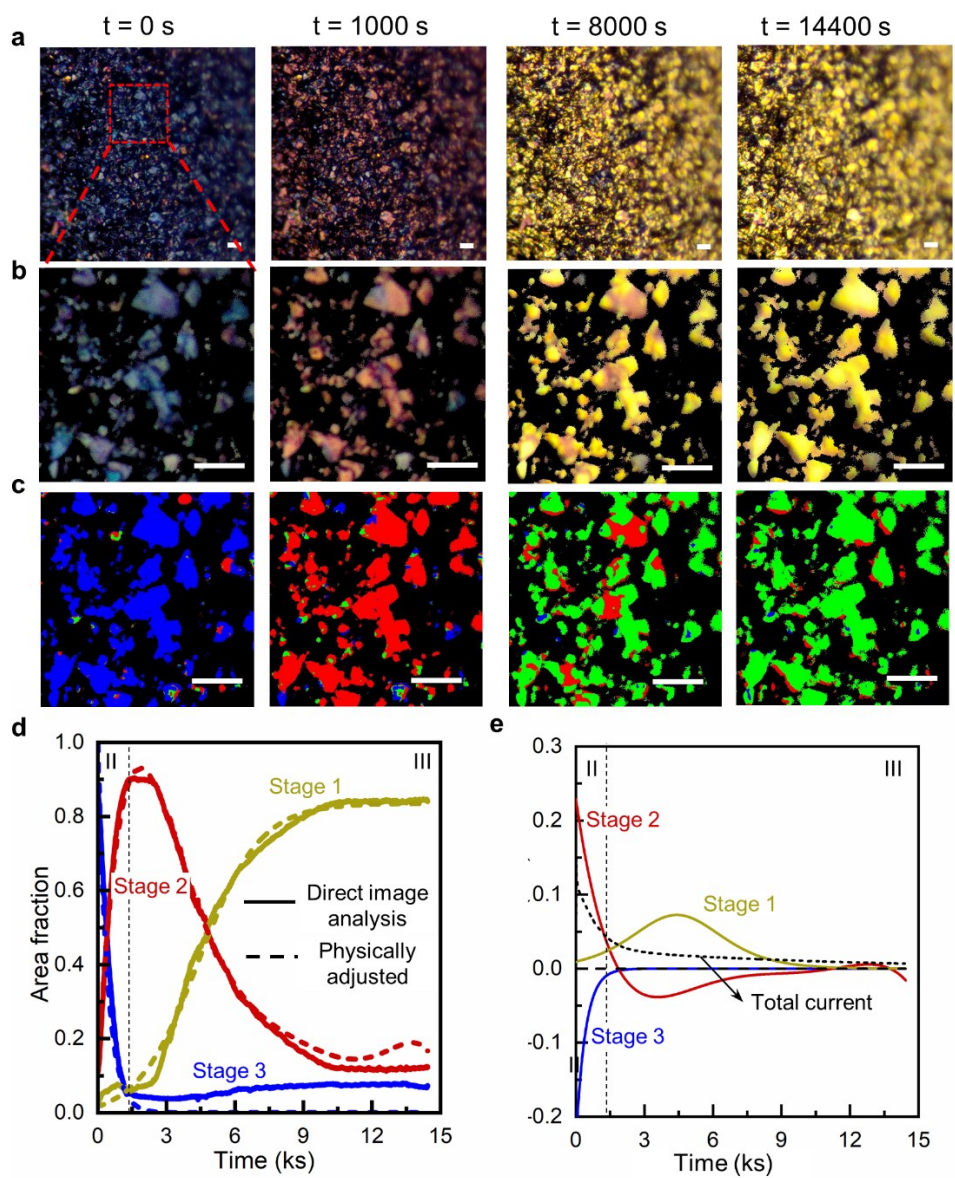

**Figure 1: Evolution of stages during Li$^+$ ion intercalation during the voltage stepping from 85 mV to 75 mV vs Li/Li$^+$. a)** Snapshots of the entire viewing area under the optical microscope with the 50x

objective, at four times: t = 0 s, 1000 s, 8000 s and 14400 s. **b)** Magnified photos highlighting the coexistence of different stages intra- and inter- particles. **c)** The converted RGB images showing the actual area fraction quantification by ImageJ. The observed blue, red and gold colors were converted to standard blue, red and green colors, respectively. **d)** The evolution curves of the colored areas during the voltage stepping, obtained from the direct image analysis along with the physically adjusted analytical curves (refer Supplementary Information Section 8 for the fitting parameters). **e)** The derived phase currents based on the time-derivative of the physically adjusted area evolution curves. *Scale bar: 10 μm*

In segment (II), a few blue particles began to turn red at the onset of the PITT voltage stepping from 85 mV to 75 mV. The localized red (Stage 2) regions always coexisted with the blue (Stage 3) regions within the same particle. The evolving boundaries between the red and blue regions clearly reveal the phase transformation process. Upon careful visual inspection, we observed that the $Li^+$ ion flux prefers to go into particles with phase boundaries. The remaining blue (Stage 3) particles will begin receiving $Li^+$ ion flux only after existing boundary-containing phase-transforming particles become completely red. The red particles then remain idle, waiting for all the other particles to reach the same stage. This process is consistent with the particle-by-particle reaction mechanism of $LiFePO_4$ electrodes at low current densities.[21] Similarly, we determine the SOC associated with Stage 2 ($x_R$) to be 55%, which is the global SOC when all particles turned red. In segment (III), while the cell is still under the same voltage held at 75 mV, the red particles start a similar particle-by-particle phase transformation process to turn gold (Stage 1). The SOC associated with Stage 1 ($x_G$) is calibrated to 100%. By converting the color into standard RGB map (Figure 1 (c)), we were able to exclude the all-time inactive region and accurately quantify the areas covered by the three colors (Stages) in thousands of *operando* snapshots. The sequential reaction process is quantitatively reflected by the evolution curves of the area fractions for the colors, shown in Figure 1 (d).

## 2.2. Currents carried by individual colors (phases)

With the SOCs for each color determined above, i.e. $x_B = 23\%$, $x_R = 55\%$. $x_G = 100\%$, the area fraction evolution curves can be converted into capacity evolution curves (**Figure S4**, Supporting Information), by $Q_i = x_i a_i(t) q_o A_T$. Here, $i$ represents Blue, Red, and Gold colors, $Q_i$ and $a_i$ re the capacity and the area fraction of color $i$. $q_o$ is the areal capacity of the entire electrode, and $A_T$ is the total area of particles accounted in the image analysis. The capacity curves directly converted from experimental data were physically adjusted based on charge conservation, to exclude possible system and sampling errors. By further taking the first order time derivative of the charge associated with each color, the phase current can be obtained,

$$I_i(t) = \frac{dQ_i}{dt} = x_i q_o A_T \left(\frac{da_i}{dt}\right) \tag{1}$$

Figure 1 (d) of the derived phase currents suggests that the stable phases grow/diminish faster than the rate of charge addition (total current), which points towards the direction that the local kinetic rate reflected by the phase boundary propagation is much higher than the electrochemical reaction rate estimated by using the total current. Since $Li^+$ ions insert into graphite particles through the edge planes, not the basal plane that reveal the colors, the area of the phase boundary, i.e. length of the phase boundary times the thickness of the particle, needs to be determined to quantify the true local current density for more accurate analysis of the kinetics. **Figure S5** in Supporting Information shows the area evolution and phase currents during 100 mV and 200 mV PITT experiments.

## 2.3. *Operando* interfacial current densities

In principle, $Li^+$ ions can intercalate into graphite particles from anywhere on the edge plane to form a shrinking-core type pattern, as observed in a 50-μm graphite disk[42] and 400-um graphite flake.[43] For our graphite particles with a mean particle size of 8.13 μm (**Figure S6**, Supporting Information), however, ion intercalations appear to occur only on a limited portion of the particle perimeter. The phase boundary, originated from the edge, appears to straighten itself to form an intercalation wave propagating through the remaining body of the particle (**Figure 2 (a)**). Since we only observe color change at the phase boundaries and not in the stable regions during Li$^+$ ion intercalation, the net flux within any color is conserved. Therefore, the flux that leads to the movement of phase boundary is identical to the reaction flux at the particle edges. Based on this observation, we propose to use the mathematical product of the total length of the phase boundaries and the thickness of the graphite particles to evaluate the true *operando* interfacial area within the porous electrodes (see Experimental Section). With proper Boolean operations (see Experimental Section and Supporting Information, **Figure S7**), the length of the phase boundaries can be determined (**Figure S8**). As shown in Figure 2 (b), the total length of the Blue-Red ($L_{BR}$) boundaries increased at the beginning of the voltage stepping from 85 mV to 75 mV, then decreased toward zero in a time span of 1250 s. While the total length of the Red-Gold boundaries ($L_{RG}$) slightly increased at the onset of voltage stepping, it remained relatively constant at a value close to zero. The trend is consistent with the decaying of the global total current. At the moment $L_{BR}$ decayed to nearly zero and stopped changing, a rapid increase in $L_{RG}$ was observed, corresponding to the onset of phase transformation from Red (Stage 2) to Gold (Stage 1), while the global total current was still decreasing, indicating a dramatic change of the true local (*operando*) interfacial current density. Similar to $L_{BR}$, $L_{RG}$ decayed after reaching its peak value but with a much slower rate in accordance with the slow decaying rate of the total current during the Red-to-Gold phase transition.

The growth periods of the three stable phases are mutually exclusive, suggesting that the entire global current at any time is carried only by one phase. Hence, by using the global current and the *operando* interfacial area calculated above, the *operando* interfacial current density, i.e., the truly working local current density, can be estimated. As shown in Figure 2 (c), the *operando* interfacial current densities are more than two orders of magnitude higher than the average current density calculated using the BET area (9.424 m$^2$·g$^{-1}$). This two-orders-of-magnitude discrepancy will be amplified to a four-orders of magnitude difference in the derived diffusion coefficients, via the square-law scaling of traditional electroanalytical methods.[23,24] This discrepancy questions the prevailing belief that the rate-limiting step of Li$^+$ ion intercalation in graphite is the bulk solid-state diffusion.

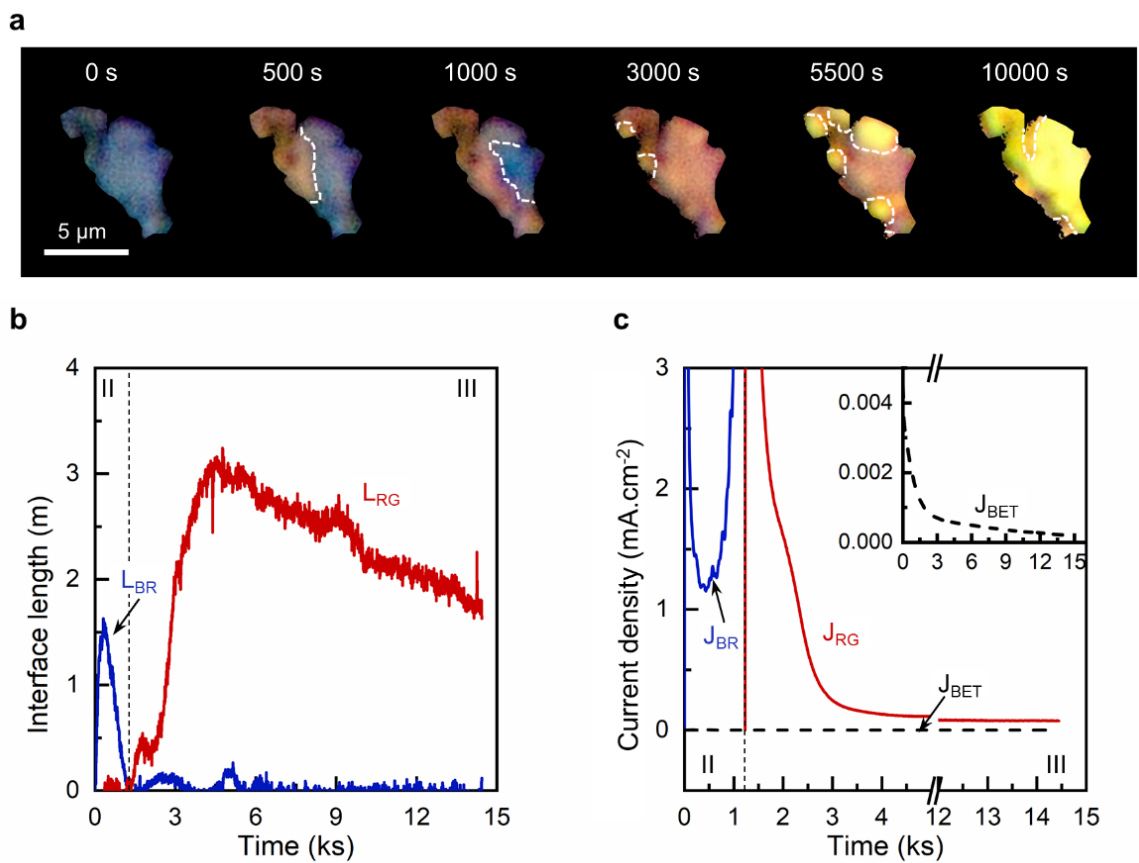

**Figure 2: Evolution of the total lengths of phase boundaries and the associated true local current densities. a)** Moving phase boundaries (white dotted lines) within a typical graphite particle during the 85 mV – 75 mV voltage stepping. **b)** Evolutions of the total lengths of interfaces between blue and red regions and between red and gold regions during the voltage step, determined from the direct image analysis of the viewing area of the electrode. **c)** The current density calculated based on the effective *operando* interfacial area. Inset shows the globally average current density based on the BET surface area. The time axis has been shrunk for improved viewing of the *operando* interfacial current density during Blue-Red transition, since it is almost constant during the Red-Gold transition.

## 2.4. Physical interpretation of the evolving phase boundaries

As suggested above, the color change only occurs at the phase boundaries during lithiation, which implies that their movement should be equivalent to the net reaction flux. At the same time, multiple nucleation and growths can emerge within a single particle (Figure 2 (a)), followed by impingements between the growing domains. The observation is consistent with the classic recrystallization process that can be analyzed by the Kolmogorov-Johnson-Mehl-Avrami (KJMA) theory.[44–48] As shown in **Figure 3 (a)** schematically, KJMA theory assumes that the ratio between the normalized *actual* incremental area of the new phase (*dA*) and the normalized "*extended*" incremental area (*dA$_{ext}$*) is always equal to the fraction of the untransformed area, *dA/dA$_{ext}$ = (1 – f)*, where *f* is the fraction of the transformed area, identical to *A*. The differential equation can be solved (see the complete derivation in Supporting Information, **Section 9**) to obtain the normalized transformed area as *A = f = 1 – exp(–A$_{ext}$)*, and further develop into the final Avrami kinetic equation by incorporating the growth rate and dimensions for the ideal $A_{ext}$. However, the challenge for electrochemical phase transformation in our graphite electrode is to quantify the evolution dynamics of the total length of the phase boundaries, instead of the area. The key question is how much of the phase boundary associated with the ideal "extended" incremental growth *dA$_{ext}$*, designated as *l$_{ext}$*, will lie in the transformed region that cannot be accounted for the total length of the *actual* phase boundary *l*, as shown in Figure 3 (a). Following the same strategy of KJMA, we propose that *l/l$_{ext}$ = (1–f) = exp(–A$_{ext}$)*. Further considering the shape factor *S*, the dimension of growth *n*, and the growth velocity *k*, we obtained the final kinetic equation for phase boundary evolution as,

$$l = nS(kt)^{n-1} \exp[-(Skt)^n] \qquad (2)$$

Interestingly, Equation (2) is essentially the first-order time derivative of the classic KJMA equation.

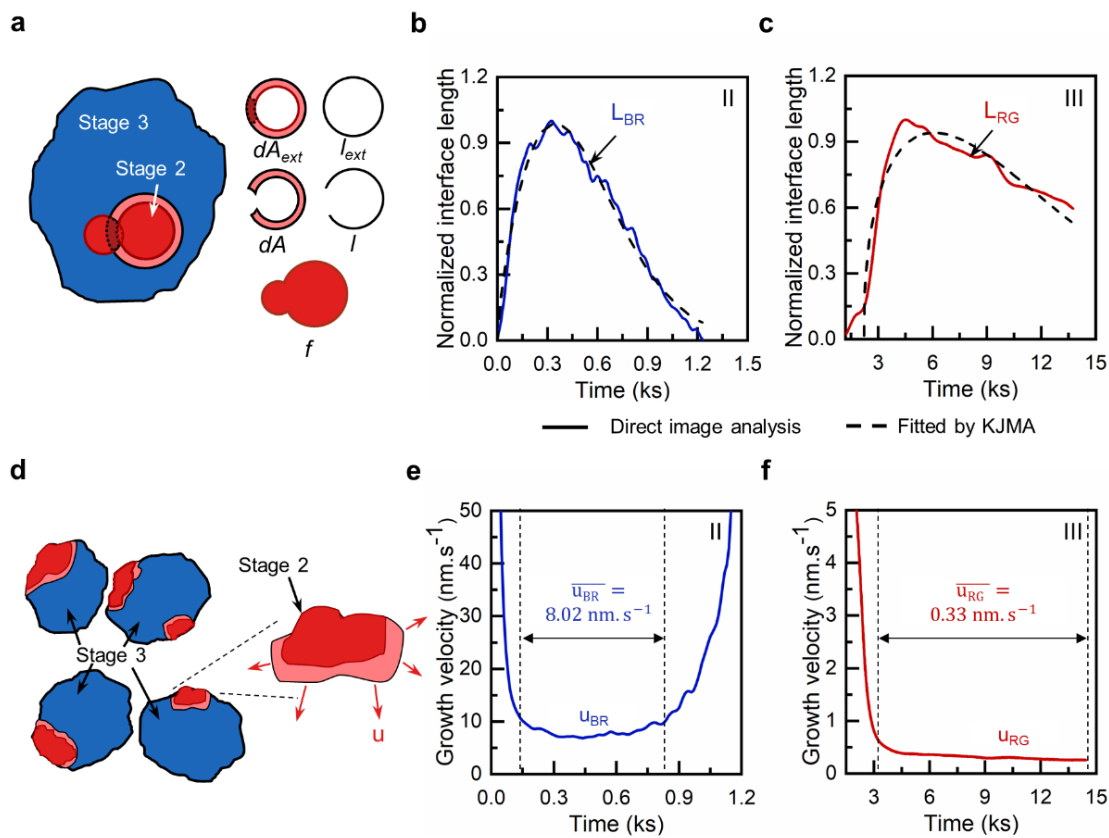

**Figure 3: Estimation of growth velocity of the stable phases using the proposed models. a)** Schematic representation of growth of Red phase onto the Blue phase explained using KJMA theory. **b)** The normalized total length of interface between blue and red regions and **c)** between red and gold regions fitted using Equation 2 to estimate the dimension of growth *n*, shape factor *S* and growth velocity *k*, during the 85 – 75mV voltage stepping. The legend on panel **c)** applies to panel **b)** as well. **d)** Schematic representation of growth of Red phase from the Blue phase in multiple particles where its growth can be tracked using the *operando* interfacial current density. **e)** The growth velocity during Blue-Red transition and **f)** Red-Gold transition, calculated from the operando interfacial current density, during the 85 – 75mV step. The marked region between the dashed vertical lines indicate the region chosen to estimate the growth velocity.

As shown in Figures 3 (b) and 3 (c), our Equation (2) fit the normalized length of phase boundary very well, especially for the boundary between the blue and red regions Figure 3 (b). While for an ideal 2-dimensional growth problem, the dimension exponent *n* should be 2, and the shape factor S should be $\sqrt{\pi} \approx$ 1.77 for an isotropic circular growth, as can be seen in the Supporting Information (Section 9), here we relaxed the constraints during the fitting to keep the analysis general. As can be seen in **Table 1**, the fitted dimension exponent $n$'s are indeed close to 2, especially for the blue-red phase boundary for all three sets of the *operando* PITT experiments (**Figure 4**), which indicate that the new phase growth through plate-like graphite particles in our thin-layer porous electrode is indeed two dimensional. The fitted shape factor $S$ is close to but not equal to the value of $\sqrt{\pi}$, indicating that the growth is not ideally isotropic as a shrinking-core process.

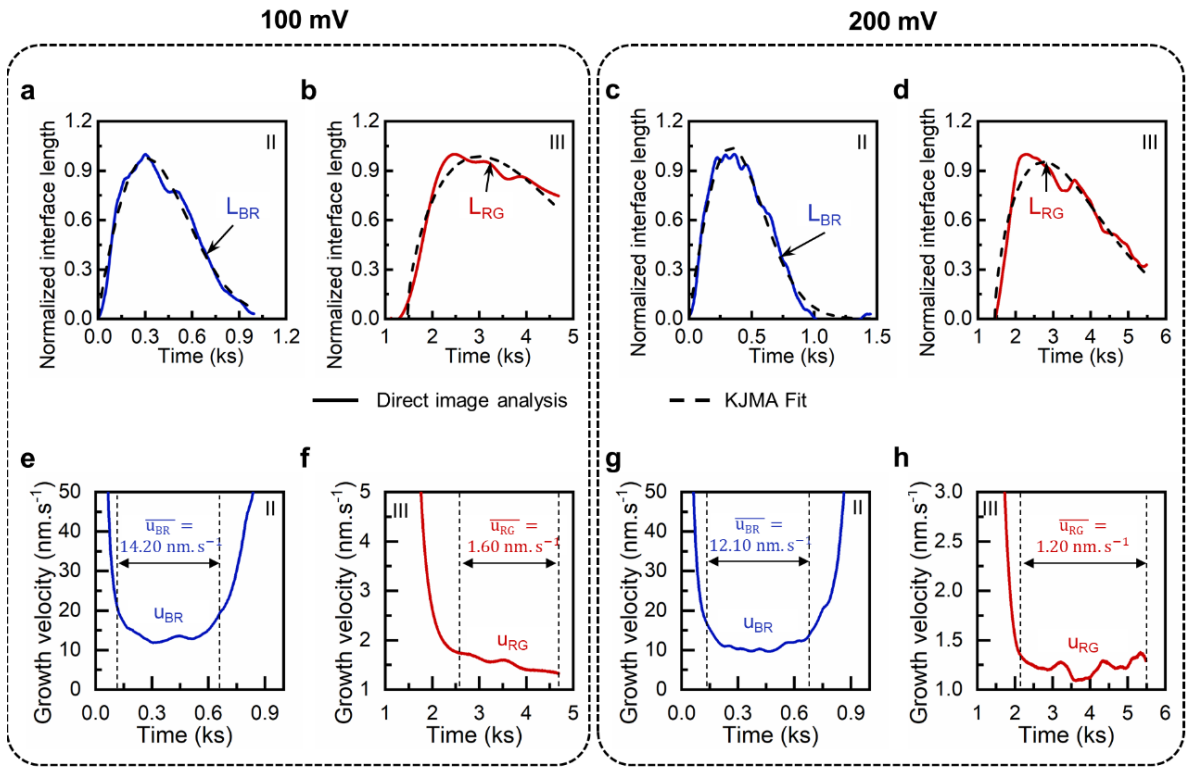

**Figure 4: Growth velocity from operando observations during phase transition induced by 100 mV and 200 mV steps.** Replication of *operando* interface length with KJMA equation during **a)** Blue-Red transition and **b)** Red-Gold transition in 100 mV case, and **c)** Blue-Red transition and **d)** Red-Gold transition in 200 mV case. Average growth velocity estimated from operando interfacial current density during **e)** Blue-Red transition and **f)** Red-Gold transition in 100 mV case, and **g)** Blue-Red transition and **h)** Red-Gold transition in 200 mV case.

The third parameter in Equation (2) also has clear physical meaning. It represents the rate of growth of the new phases, i.e. the propagation velocity of the phase boundary, which is a direct result of the net reaction flux and can be quantified from the *operando* experiments by using mass conservation.

$$u(t) = j^{int} \frac{V_m}{F} \tag{3}$$

$u(t)$ is the *effective* one-dimensional growth velocity of the new phase at time $t$, $j^{int}$ is the interfacial current density of the corresponding phase calculated by dividing the global current $I(t)$ with the *operando* interface area $A(t)$. $V_m$ is the molar volume of that particular phase of the lithiated graphite, and $F$ is the Faraday constant. Figure 3 (d) schematically shows the growth of the new phase. The complete derivation can be found in the Supporting Information, **Section 10**).

Equation (3) shows that the growth velocity can be directly estimated from the *operando* interfacial current density. Since the *operando* interfacial current densities are relatively stable for the major portion of the corresponding phase transformation segments, despite spikes at the beginning and/or the end, an averaged growth velocity ($\bar{u}$) can be calculated (**Figures 3 (e-f), and Figures 4 (e-h)**). Table 1 lists the fitting parameters obtained from Equation (2) for all three cases, along with the averaged growth velocities obtained by using *operando* interfacial current densities. It is noteworthy that Equation (2) returns the growth velocity $k$ with the units of s$^{-1}$. To find the actual growth velocity, we solved for Equation (2) with actual dimensions shown in the Supporting Information, Section 9. Equation (2) that fits the transient total length of the phase boundary, and Equation (3) that derived from the ratio between the transient total current and the transient phase boundary, independently explain the same phenomena from different perspective, but surprisingly obtain almost the same growth (i.e. interface propagation) velocity in the units of nm·s$^{-1}$. The self-consistent results not only validate our new model on the kinetics of the phase boundary evolution, but also indicate that the dynamic is controlled by the *operando* (i.e. local and working) electrochemical reaction flux, not solid-state diffusion.

**Table 1: Parameters obtained from fittings the *operando* interface lengths with Equation (2) and the averaged growth velocity calculated from *operando* interfacial current densities**

|  |  | 10 mV | 100mV | 200 mV |
|---|---|---|---|---|
| **Blue-Red** | n | 1.75 | 1.90 | 1.98 |
|  | S | 1.48 | 1.36 | 1.48 |
|  | $k_{fit}$ (nm·s$^{-1}$) | 13.4 | 11.6 | 15.8 |
|  | $k_{exp} = \bar{u}$ (nm·s$^{-1}$) | 8.02 | 14.20 | 12.10 |
| **Red-Gold** | n | 1.39 | 1.58 | 1.60 |
|  | S | 1.41 | 1.52 | 1.43 |
|  | $k_{fit}$ (nm·s$^{-1}$) | 0.5 | 1.5 | 2.2 |
|  | $k_{exp} = \bar{u}$ (nm·s$^{-1}$) | 0.3 | 1.6 | 1.2 |

## 2.5. Determination of the diffusion coefficients

The direct correlation between the growth velocity and the *operando* interfacial current density revealed above raises the question whether the solid-state diffusion coefficient can still be reliably extracted from the standard PITT experiment originally established specifically for the evaluation of solid-state diffusion.[49] As can be seen in **Figures 5 (a-c)**, the lack of straight lines (predicted by the classic Cottrell equation) suggests that the processes are likely not diffusion-limited. As our first attempt, we adopted the modified PITT (mPITT) model,[50,51] without the presumption of the rate-limiting step, to fit the *operando* interfacial current densities. Both the diffusion coefficient $D_{Li}$ and the electrochemical Biot number $B$, can be obtained by minimizing the least squares. Good agreements can be found for the solid solution processes (**Figure 5 (a)**), but not for the Blue-Red phase transformation (**Figure 5(b)**). Interestingly, the mPITT model (Equation (S7a) in Supporting Information **Section 11**) can fit the *operando* interfacial current density for the Red-Gold phase transformation in the $t << l^2/D_{Li}$ regime fairly well, especially in the PITT experiments with 10 mV and 200 mV steps. It's worth noting that, among all the recorded phase transformation processes, only the Red to Gold transformation was long enough to enter in the regime of $t >> l^2/D_{Li}$, but cannot be fitted by the mPITT model for $t >> l^2/D_{Li}$, i.e. Equation (S7b) in Supporting

Information Section 11, as the *operando* interfacial current densities remain relatively constant. Although the mPITT model does not specifically take into account the phase transformation processes, it holds the generality from its origin of direct mathematical approximations to the battery voltage and current. Therefore, the fittings for the Red-Gold phase transformation process may still provide insights from the apparent good agreements. A critical feature of the mPITT model is that the apparent two independent fitting parameters are actually constrained by the system-specific exchange current density ($j_0$) via the definition of Biot number, $B=-j_0 l(\partial U/\partial C)/(D_{Li}RT)$. Here, $\partial U/\partial C$ is the derivative of the open-circuit voltage (OCV) with respect to the solid-state Li$^+$ ion concentration, $R$ is the gas constant and $T$ is the temperature. Again, the model becomes inapplicable in the "ideal" phase-transformation regimes due to $\partial U/\partial C$ being zero leaving $j_0$ impossible to identify. Nevertheless, the application of a modified PITT (mPITT) model[50,51] without presumption of the rate-limiting step can give a rough estimate of the diffusion coefficients of Li$^+$ ions into graphite. The Li$^+$ ion diffusion coefficients, extracted from fitting the mPITT model on the *operando* interfacial current density during the solid-solution lithiation (Segment (I)), lies between $1.35 \times 10^{-11} - 3.27 \times 10^{-10}$ cm$^2 \cdot$s$^{-1}$, consistent with the reported values.[28–31,33,34] The corresponding electrochemical Biot numbers ($B$) shown in Figure 5 (d) suggest a diffusion-limited process. On the other hand, the diffusion coefficients for the Red-Gold phase transformation processes lying in the range $1.98 \times 10^{-8} - 1.31 \times 10^{-7}$ cm$^2 \cdot$s$^{-1}$ across the three sets of experiments, which are in very good agreements not only with the Cahn-Hilliard phase field simulation of a 50-um graphite disk,[42] but also the first principles calculations.[37] Such a fast diffusion means a very low diffusion time constant, $\tau_D = L^2/D_{Li} = 1 - 8$ s where $L$ is the diffusion length, set to be one half of the dimension of basal plane,[29] i.e. 4 μm for our case. **Figure S9** in the Supporting Information shows consistent interpretation for the responses in other PITT voltage steps during solid-solution intercalation. Our results provide a straightforward evidence that once the spatiotemporal heterogeneities emerge, the process is no longer diffusion-limited, consistent with a recent scaling analysis.[43]

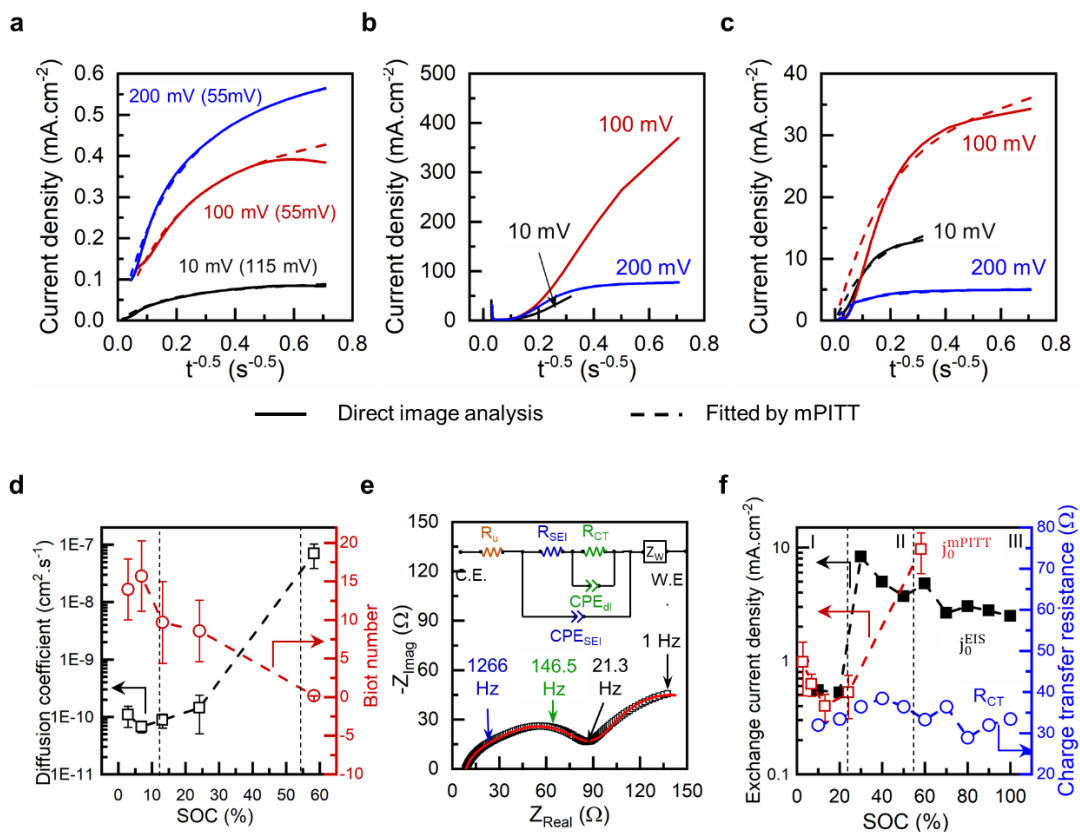

**Figure 5: Independent experiments to estimate the kinetic parameters. a)** The fitting of representative *operando* interfacial current density during solid-solution processes in all the three sets of experiments with the modified PITT model to extract the Li$^+$ ion diffusion coefficient into graphite. The voltage mentioned in brackets are the voltage step during the corresponding PITT experiment. **b)** The *operando* interfacial current density during Blue-Red phase transformation in all the three sets of experiments. They could not be fitted with the modified PITT model owing to their highly non-monotonic nature. **c)** The fitting of *operando* interfacial current density during Red-Gold phase transformation in all the three sets of experiments with the modified PITT model to extract the Li$^+$ ion diffusion coefficient into graphite. **d)** Extracted $D_{Li}$ and $B$ from the fitting the operando interfacial current densities in panel (a) and (c) with the mPITT model. The vertical dashed lines separate the entire region into diffusion-controlled, mixed control and reaction-controlled regimes. **e)** Fitting of experimental EIS data with the specified equivalent circuit model composed of two RC circuits and a porous-bounded Warburg impedance. The blue and red semi-circles are the individual contribution of the two RC circuits with the frequencies of the respective peaks being $\omega_{SEI}$ and $\omega_{dl}$. Here,

$R_{SEI}$ and $R_{CT}$ are the charge-transfer resistances of the SEI and double layer, $CPE_{SEI}$ and $CPE_{dl}$ are the constant phase elements of SEI and double layer, and $Z_W$ is the porous-bounded Warburg impedance. **f)** Exchange current density estimated from the EIS using *operando* interface area and that from mPITT averaged over three sets of experiments along with its close agreement with the exchange current densities calculated from $D_{Li}$ and $B$ using mPITT model.

## 2.6. Impedance analysis for the *operando* exchange current densities

While the mPITT method itself allows the evaluation of $j_0$, independent electrochemical impedance spectroscopy (EIS) experiments were performed on the same *operando* cells to assess the interfacial reaction kinetics during Li$^+$ ion intercalation. **Figure 5 (e)** shows the Nyquist plot of a typical EIS spectra, which appears to have two partially merged semicircles followed by a Warburg tail. While the Warburg tail can be attributed to the solid diffusion in graphite particles, the two semicircles need careful examination of their physical basis. Based on the consensus that the charge transfer reaction occurs on the surface of graphite particles, but beneath the solid electrolyte interphase (SEI) layer, we hypothesize that (i) the electrical double layer capacitance associated with the charge transfer will form within the SEI; and (ii) the charges at the electrolyte|SEI interface and those at the SEI|graphite interface will form a double-plate capacitor filled with the SEI layer. To test the hypothesis, the EIS spectra was fitted with the equivalent circuit model shown in **Figure 5(e)** to obtain the two resistances ($R_{SEI}$ = 38.4 Ω and $R_{CT}$ = 32.0 Ω), which were then use to calculate the capacitances from the two characteristic frequencies ($\omega_{SEI}$ = 1266 Hz and $\omega_{CT}$ = 146.5 Hz) labeled in Figure 5 (e), via the general formula $\omega = (RC)^{-1}$. The calculated capacitances ($C_{SEI}$ = 20.5 μF and $C_{CT}$ = 21.4 μF) were then used to determine the associated dielectric constants via $C = \varepsilon_r \varepsilon_0 A_{edge}/d$, where $A_{edge}$ is the total area of the edge planes, $d$ is the thickness of the respective capacitors (assuming 10 nm for the SEI layer and 1 nm for the electrical double layer), $\varepsilon_0$ is the absolute permittivity of vacuum and $\varepsilon_r$ is the dielectric constant. The resulting dielectric constants (161.2 and 167.5) are surprisingly very close to each other, but much higher than the dielectric constants for liquid electrolyte (<90) we used in the operando cells,[52] yet consistent with the reported values for the solid electrolyte interphase[53] and solid polymer electrolytes.[54] This quantitative physical confirmation validates our hypothesis on the microstructure of the equivalent circuit model. All the parameters obtained from the fittings of the equivalent circuit model at 10% - 100% SOCs are displayed in Table S3 in Supporting Information **section 12**. The charge-transfer resistance ($R_{CT}$) thus obtained were used to calculate both the exchange current density via[51] $j_0 = RT/(FR_{CT} A_{int})$ as shown in Figure 5 (f), and the reaction time constant $\tau_R = Q/j_0 A_{int} = (QFR_{CT})/(RT)$. Here, $A_{int}$ is again the *operando* interfacial area, $Q$ is the total charge transferred to the electrode, $R$ is the gas constant and $F$ is the Faraday constant. These values of $j_0$ are in good agreement with those obtained from mPITT fitting at selected SOCs. The reaction time constant $\tau_R$ calculated from the maximum capacity of the electrode lies between 488 and 680 s for the three sets of experiments. These maximum possible values of $\tau_R$ is almost half of the Blue-Red transition time (1000 – 1250 s), and 7 – 20 times lower than the Red-Gold transition time (5000 – 13000 s). While this comparison between time constants seems to suggest that the process is not limited by reaction, it is the actual *operando* interfacial current density, not the exchange current density or the exchange reaction rate that plays the role in the actual process, which was also pointed out by Fraggedakis et al.[43]

## 3. Discussion

From analyzing the dynamics of phase boundary evolution by using Equation (2) and evaluating the true local exchange current density by using the physics-based equivalent circuit model, our results clearly suggest that the electrochemical lithiation process of particulate graphite electrode is not diffusion-limited. Therefore, classic electroanalytical methods based on the diffusion-limited assumption are not appropriate for determining the diffusion coefficient, even if the true local current density can be accurately determined. This is particularly important for phase-transformation electrodes, for which the classic methods require the term $dU/dC$ that becomes zero at the phase-transformation plateau, making other physical quantities impossible to identify. This is also the reason that the Warburg fittings from our physics-based EIS analysis cannot be used to determine the diffusion coefficient, as the $dU/dC$ term is again required for the Huggins equation to convert the Warburg factor.[55] From a deeper perspective, the original derivation[55] of now the widely used Warburg impedance did not take into account any phase transformation processes. Even if one can get around the *dU/dC* term,[56] the obtained the diffusion coefficients are still several orders of magnitude lower (**Table S5** in Supporting Information), by which a Cottrell-type diffusion-limited process is dictated but does not exist in all our experiments. The only reliable electrochemical method of obtaining the diffusion coefficient in phase transformation materials appears to be modeling the transients from a single particle with precise *operando* current density.[42]

Regarding the rate-limiting step of electrochemical processes in particulate phase transformation electrodes, our results rejects the simple yet convenient terms of diffusion-limited or reaction-limited. It should rather be designated as phase-transformation controlled for two reasons at two different scales. First, at the single particle scale, the reaction rate is controlled externally by electrochemical driving force. Therefore, whether the entire process is diffusion-limited or reaction-limited is extrinsic. Unless the local current density for the single particle can be exclusively and precisely determined and controlled, determining the rate-limiting step and the kinetic parameters would remain very challenging. Second, at the mesoscale with at least hundreds of particles, while one can precisely determine and control the *total* current, the self-adapted local driving force would activate different numbers of particles[48] to share the total current, leading to unexpected *operando* (local and working) current density. As can be seen in **Figure 5 (b) and (c)**, the *operando* interfacial current densities from 200 mV PITT experiments are surprisingly and significantly lower than those from 10 mV and 100 mV PITT experiments, which are the exact results of that the higher driving force (200 mV versus 100 mV and 10 mV) promoted more phase transforming particles with longer total phase boundaries. It is the phase transformation dynamics that controls how many particles and how much interfacial area will be activated for working, and it is still the phase transformation dynamics that controls the evolution of phase boundaries within each particle. Therefore, it should be recognized that the phase-transformation control mechanisms at both scales induce the spatiotemporal heterogeneities and limit the overall performance of the electrode.

Our results also raise a fundamental question: whether the electrochemical response from electrode under small excitations (e.g. low total current) can be considered from quasi- or near-equilibrium physical processes. It is apparent now that the electrochemical responses of the electrode (total current and terminal voltage) are collective behaviors of *far-from-equilibrium* dynamics contributed only from a small portion of the electrodes. The insights from this study stress the necessity of careful examination of the local electrochemical activities.

Our method is not limited to the PITT technique. Revealing the mesoscale spatiotemporal heterogeneities under other types of electrochemical excitations (e.g. galvanostatic cycling, cyclic voltammetry, etc.) will lead to critical refinements to existing understandings of the electrochemical kinetics and rate-limiting steps. For electrode materials without this unique visible color-changing property, Raman spectroscopy appears to an effective monitoring approach to achieve the heterogeneity map,[57] which can be further analyzed following our methods to obtain the *operando* interfacial kinetics.

## 4. Conclusion

By exploiting the colorimetric behavior of lithiated graphite, we have demonstrated a direct, simple, yet precision method to monitor and quantify the spatiotemporal heterogeneities in particulate porous electrodes. The true local current density, i.e. the *operando* interfacial current density by our definition, obtained from direct image analysis is ~100 times higher than the BET-average current density. Although all the particles in the porous electrodes are electrochemically active and eventually get fully intercalated with the $Li^+$ ions, at any time instant, only a limited number of particles and limited portion of the total available area receive the ionic flux. Our *operando* monitoring clearly revealed that once a successful nucleation event occurs and phase boundaries start to form in a randomly chosen particle, it is preferred for further intercalation irrespective of its shape and size. Since the $Li^+$ intercalation into graphite particles is not diffusion-limited, smaller particles do not necessarily provide a substantial improvement in the high-rate performance. However, reducing particle size may help eliminate the reaction heterogeneities by altering the nucleation barrier for solid-state phase transformation in individual particles, such that the concurrent reaction pathway become thermodynamically favorable.

## 5. Experimental Section

**Thin electrode preparation:** Graphite flakes (7-10 μm, 99%, Alfa-Aesar), PVdF binder (>99.5%, MTI Corp) and conductive acetylene black powder (35-40 nm, MTI Corp) were mixed in the ratio 88:10:2 and dissolved in 1-methyl-2-pyrrolidone (NMP, 99.5%, Sigma Aldrich) to form a homogeneous slurry. To ensure the best imaging quality, the slurry was coated onto separator film by the doctor-blade method. The electrodes were dried at 60°C to remove the NMP. Φ8 mm electrodes were punched out and were kept under vacuum at 70°C for 12 hours to remove the residual moisture. The active material loading, and electrode thickness were 0.7 mg cm$^{-2}$ and 5 μm, respectively. The SEM images of the electrode are shown in the Figure S1, Supporting Information.

*Operando* **setup and experiments.** A half-cell using the thin graphite electrode, a Li anode, a glass-fiber separator, and 1M LiPF$_6$ in EC:DMC (50:50 v/v) in a standard 2032 coin cell with a 2 mm hole on the top, was assembled in an Ar-filled glovebox. A 5 x 5 mm glass window was attached using epoxy to seal the cell and view the graphite flakes under the optical microscope. The coin cell was placed on a stage of the Olympus BX53M microscope under objective 50x for *operando* observation. The cell was cycled at C/4 current five times between cut-off voltages 1.5V and 0.4 mV, to form a stable SEI. We then, performed a three PITT discharge experiments from 245 mV to 0.1 mV, varying the voltage steps and C/20 threshold current: 1), with 10 mV, 2) 100 mV, and 3) 200 mV step sizes, while capturing the time frames every 10s. All the acquired digital photos were processed using ImageJ to quantify the colored regions. The detailed description of the procedures is mentioned below. See sections 4-7 of the Supporting Information for more details.

An Electrochemical Impedance Spectroscopy (EIS) was conducted at intervals of 10% SOC from 10% to 100% SOCs. The cells were discharged at 0.1C current to the relevant SOC and relaxed for 2h to reach equilibrium before taking the EIS measurements. All the EIS measurements were taken at 10 mV amplitude in the frequency range 1 MHz – 1 Hz. The obtained Nyquist plots were fitted using the equivalent circuit model, shown in Figure 5 (a).

**Color thresholds for area quantification.** The built-in Hue-Saturation-Brightness threshold method of ImageJ was used to identify the blue, red and gold colors in the photos captured in the *operando* PITT experiment. ImageJ auto-selects the brightness to accommodate all the non-black regions/graphite flakes. All the grey/unreacted regions were first selected between the saturation range 0 – 40 and were converted to black. A fixed range of hue was used to select similar colored regions (Red: 0 – 24, Gold: 24 – 44, and Blue: 44 – 255) while maintaining the same range of saturation (40 – 255) and brightness. The leftover regions or the surrounding matrix, calibrated out at 100% SOC, were converted to the standard black color. The above criteria were applied to all the photos with the help of a script.

**Charge conservation calibration.** The area evolution curve in Figure 1 (c) was investigated to understand how the stable phases change, which is responsible for the surface reaction, by applying the charge conservation within the electrode, $\sum_i q_i A_i(t) = q_T(t) A_T$, where $i$ represents Blue, Red, and Gold, $q_i$ is the areal capacity of the $i^{th}$ color and can be calculated from the SOC of the $i^{th}$ color (estimated above) and the theoretical areal capacity of the material ($q_o$), $q_i = x_i q_o$, $A_i$ is the area covered by $i^{th}$ color, $q_T$ is the areal capacity of the electrode and $A_T$ is the total surface area of particles in the electrode. The above equation can, then be transformed into $\sum_i x_i a_i(t) = x_T(t)$ where $a_i$ is the area fraction of the $i^{th}$ color, and $x_T$ is the global SOC of the electrode. For known area fractions of stable phases, a phase-transforming material should inherently follow this equation. Since we only observed a $100 \times 100\mu m$ window under the optical microscope, the validity of the above equation for the observed region confirms that the analysis can be confidently extrapolated to the entire electrode (Supporting Information, Section 8).

**Curve validation and physical adjustment.** The capacity carried by each stage during the PITT discharge was calculated using $Q_i = q_i A_i(t) = x_i q_o a_i(t) A_T$, as explained in the main text. The equation is valid because the intercalation process satisfied the equation $\sum_i x_i a_i(t) = x_T(t)$ during the entire PITT (See Supporting Information Section S8). The individual capacity contribution follows the same trend as the area fraction evolution curve in Figure 1 (c). They were individually represented by analytical expressions, for instance, Stage 3 by an exponential curve, Stage 2 by a 6th order polynomial equation and Stage 1 by a logistic S-shaped curve. A detailed description with fitting parameters is provided in the Supporting Information Section S8. The phase currents were calculated by taking a 1st order time derivative of the obtained analytical expressions (Equation 1).

Small noises were observed while estimating the length of interface using ImageJ (Figure 2 (a)) due to errors arising in the pixel-by-pixel measurement. We removed this noise by applying a quadratic regression method in MATLAB, enabling us to obtain smooth interfacial current densities.

**Determination of Interface Area:** For a shape with two colors, ImageJ can be used to find the perimeter covered by each color, and the outer perimeter of the shape, which together can be solved for the length of the interface. In our case, particles existed in three different states at a time. To calculate the length of the interface, for instance, the Blue – Red interface, we relied on the fact that the phase transformation in graphite can only occur in one order: Stage 3 to Stage 2 to Stage 1. We converted all the green regions in the transformed RGB images, to the standard red color, thus eliminating all Red – Gold interfaces. This enabled us to find the length of the Blue – Red interface using the above methodology from the following equation, $L_{BR} = (l_{Blue} + l_{Red} - l_{particles})/2$. Similarly, we calculated the length of the Red – Gold interface by converting all the blue regions to the standard red and applying following equation, $L_{RG} = (l_{Red} + l_{Gold} - l_{particles})/2$ where $L_{BR}$ and $L_{RG}$ are the lengths of Blue – Red and Red – Gold interfaces respectively, $l_{Blue}$, $l_{Red}$ and $l_{Gold}$ are the perimeters of the blue, red and gold regions in the corresponding transformed images, and $l_{particles}$ is the outer perimeter of all the particles within the viewing frame.

Considering disc-shaped flakes with an average diameter 8μm and thickness 0.5 μm, our ~5 μm thick electrode constituted ~$10^7$ particles with 10 layers stacked over each other. On making a statistical assumption that all layers were similar, we calculated the active area by multiplying the total length of interface with the electrode thickness.

**Acknowledgement**

P.B. acknowledges the startup support from Washington University in St. Louis. The materials characterization experiments were partially supported by IMSE (Institute of Materials Science and Engineering) and by a grant from InCEES (International Center for Energy, Environment and Sustainability) at Washington University in Saint Louis.

**Author contributions**

P.B. conceived and supervised the study. S.A. and P.B. designed the experiments. S.A. performed the experiments, carried out the analysis. S.A and P.B. wrote and revised the manuscript.

**Additional information**

Supplementary information is available online.

**Competing interests**

The authors declare no competing financial interests.

**Data availability**

The data that support the findings of this study are available from the corresponding author upon reasonable request.

**Code availability**

The ImageJ scripts for direct image analysis and MATLAB code for model calculation are available from the corresponding author upon reasonable request.